\documentclass[journal,twoside,web]{ieeecolor}
\usepackage{generic}
\usepackage{cite}
\usepackage{amsmath,amssymb,amsfonts}
\usepackage{multirow}
\usepackage{graphicx}
\usepackage{textcomp}
\usepackage{tabularx}
\usepackage{xcolor}
\usepackage{soul,dsfont} 
\usepackage{tikz} 
\usepackage{color}
\usepackage{algorithm,algorithmicx}
\usepackage{array}
\usepackage{eqparbox}
\usepackage{url}
\usepackage{relsize}
\usepackage{stfloats}
\usepackage{graphics}
\usepackage[mathscr]{euscript}
\graphicspath{{figures/}{./}}
\usepackage{algpseudocode}

\newtheorem{definition}{Definition}

\newtheorem{remark}{Remark}

\def\BibTeX{{\rm B\kern-.05em{\sc i\kern-.025em b}\kern-.08em
    T\kern-.1667em\lower.7ex\hbox{E}\kern-.125emX}}
\begin{document}

\bstctlcite{}
\title{Tensor Dynamic Mode Decomposition}
\author{Ziqin He, 
Mengqi Hu,
Yifei Lou, and 
Can Chen
\thanks{Ziqin He and Mengqi Hu are with the Department of Mathematics,  University of North Carolina at Chapel Hill, Chapel Hill, NC 27599, USA (email: zhe21@unc.edu; qmhu123123@gmail.com).}
\thanks{Yifei Lou is with the Department of Mathematics, and the School of Data Science and Society,  University of North Carolina at Chapel Hill, Chapel Hill, NC 27599, USA (email: yflou@unc.edu).}
\thanks{Can Chen is with the School of Data Science and Society, the Department of Mathematics, and the Department of Biostatistics,  University of North Carolina at Chapel Hill, Chapel Hill, NC 27599, USA (email: canc@unc.edu).}
}

\maketitle
\begin{abstract}
Dynamic mode decomposition (DMD) has become a powerful data-driven method for analyzing the spatiotemporal dynamics of complex, high-dimensional systems. However, conventional DMD methods are limited to matrix-based formulations, which might be inefficient or inadequate for modeling inherently multidimensional data including images, videos, and higher-order networks. In this letter, we propose tensor dynamic mode decomposition (TDMD), a novel extension of DMD to third-order tensors based on the recently developed T-product framework. By incorporating tensor factorization techniques, TDMD achieves more efficient computation and better preservation of spatial and temporal structures in multiway data for tasks such as state reconstruction and dynamic component separation, compared to standard DMD with data flattening. We demonstrate the effectiveness of TDMD on both synthetic and real-world datasets.   
\end{abstract}

\section{Introduction}
Dynamic mode decomposition (DMD) has emerged as a powerful data-driven technique for analyzing the dynamics of complex systems from high-dimensional time-series data \cite{schmid2022dynamic,tu2013dynamic,proctor2016dynamic}. Originally developed in fluid dynamics \cite{schmid2010dynamic}, DMD decomposes a sequence of system snapshots into spatial modes and corresponding temporal dynamics, capturing coherent structures and their evolution over time. DMD has been successfully applied across a wide range of domains, including video background modeling \cite{kutz2017dynamic}, neural data analysis \cite{brunton2016extracting}, robotics \cite{berger2015estimation}, power systems \cite{alassaf2019dynamic}, and epidemiology \cite{proctor2015discovering}. Its ability to extract low-rank models and reveal dominant dynamical patterns has led to widespread adoption in both scientific research and real-world engineering systems. However, most existing DMD methods operate on vector-based data and are not directly applicable to inherently multidimensional data such as images, videos, and higher-order networks.

Tensors are multidimensional arrays that extend vectors and matrices \textcolor{black}{to higher dimensions}, providing a more natural and accurate representation of multiway data \cite{kolda2009tensor}. A variety of tensor-based methods have been developed to analyze and control  dynamical systems involving multiway data \cite{chen2021multilinear}. In particular, Klus et al. \cite{klus2018tensor}  proposed a tensor-based DMD approach that  leverages  tensor train decomposition to compute DMD modes and eigenvalues directly from high-dimensional data, reducing computational and memory requirements for high-dimensional data.
Additionally, Zhang et al. \cite{10706099} introduced high-order dynamic mode decomposition, a hybrid method that combines high-order singular value decomposition with DMD to improve robustness against noise in multidimensional harmonic retrieval.
\textcolor{black}{While both approaches incorporate tensor decomposition techniques, their core remains rooted in standard DMD (the former simply reformulates DMD using tensor train representations, while the latter embeds DMD within a hybrid framework),  rather than fundamentally transforming the DMD framework through tensor-based formulations.}

To address this, we aim to leverage tensor-based dynamical systems and tensor algebra to extend DMD into higher-order tensors. Our  focus in this work is on third-order tensors, which frequently arise in applications such as images, videos, and spatiotemporal sensor arrays where data is indexed along two spatial dimensions and one temporal or feature dimension \cite{kilmer2013third,braman2010third}. To effectively model and manipulate such data, we adopt the T-product, a specialized tensor multiplication operation defined for third-order tensors \cite{lund2020tensor}. The T-product generalizes fundamental linear algebraic operations to the tensor domain, enabling a matrix-like algebra that preserves the inherent multiway structure and supports the development of efficient tensor-based algorithms for dynamical system analysis \cite{10848219,he2025data}. Notably, tensor factorization techniques such as tensor singular value decomposition (TSVD) and tensor eigenvalue decomposition (TEVD) \cite{kilmer2011factorization} can be employed to extract low-rank tensor structures, identify dominant dynamical modes, and reduce computational complexity.

In this letter, we propose tensor dynamic mode decomposition (TDMD), a novel extension of DMD to third-order tensors. TDMD employs economy-size or truncated TSVD on the observed state data to construct a reduced-order transition tensor, followed by TEVD on this reduced tensor to extract the corresponding TDMD modes. We demonstrate that TDMD can effectively perform  tasks such as state reconstruction and dynamic component separation. The framework is validated on both synthetic and real-world video data, highlighting its advantages over standard DMD. \textcolor{black}{Notably, TDMD can be  extended to higher-order tensors through the generalized T-product \cite{cates2022anomaly}.} By operating directly on multidimensional data without flattening, TDMD better preserves spatial-temporal structures and reduces memory and computational costs, making it particularly promising for large-scale applications in video processing, biomedical imaging, and sensor networks.

The rest of this letter is structured as follows. In Section \ref{sec:prelim}, we provide an overview of T-product operations and a brief review of DMD. In Section \ref{sec:tdmd}, we present the detailed procedure of the proposed TDMD method and demonstrate its applications in state reconstruction and dynamic component separation.   In Section \ref{sec:num}, we offer numerical experiments to illustrate the performance of TDMD.  Finally, we conclude with future directions
in Section \ref{sec:con}.

\section{Preliminaries}\label{sec:prelim}

\subsection{T-Product Algebra}
We focus specifically on third-order tensors, which are multidimensional arrays with three modes or axes, and adapt the T-product framework \cite{lund2020tensor,chen2024tensor} to facilitate computations within this tensor space. Given two third-order tensors $\mathscr{X}\in\mathbb{R}^{n\times h\times m}$ and $\mathscr{Y}\in\mathbb{R}^{h\times s\times m}$, the T-product is defined as
\begin{equation}
\mathscr{X}\star\mathscr{Y} = \texttt{fold}\big(\texttt{bcirc}(\mathscr{X})\texttt{unfold}(\mathscr{Y})\big) \in\mathbb{R}^{n\times s\times m},
\end{equation}
where the operations $\texttt{bcirc}$ and $\texttt{unfold}$ are defined as
\begin{align*}\label{eq:bcirc}
\texttt{bcirc}(\mathscr{X})&= \begin{bmatrix}
\mathscr{X}_{::1} & \mathscr{X}_{::m} & \cdots & \mathscr{X}_{::2}\\
\mathscr{X}_{::2} & \mathscr{X}_{::1} & \cdots & \mathscr{X}_{::3}\\
\vdots & \vdots & \ddots & \vdots\\
\mathscr{X}_{::m} & \mathscr{X}_{::(m-1)} & \cdots & \mathscr{X}_{::1}
\end{bmatrix}\in\mathbb{R}^{nm\times hm},\\
\texttt{unfold}(\mathscr{Y}) &= \begin{bmatrix}
\mathscr{Y}_{::1}^\top &
\mathscr{Y}_{::2}^\top&
\cdots&
\mathscr{Y}_{::m}^\top
\end{bmatrix}^\top\in\mathbb{R}^{hm\times s},
\end{align*}
and \texttt{fold} is the  reverse operation of \texttt{unfold}. Here, the notation : refers to the MATLAB colon operator, which acts as a shorthand to indicate all indices along a given mode of the tensor, and  $\mathscr{X}_{::j}$ denotes the $j$th frontal slice of  $\mathscr{X}$, which is a matrix slice along the third mode.

Several  linear algebraic operations can be  extended to third-order tensors through the T-product framework. The T-identity tensor $\mathscr{I} \in \mathbb{R}^{n \times n \times m}$ is defined such that its first frontal slice $\mathscr{I}_{::1}$ is the identity matrix, while all other frontal slices are zero matrices. The T-transpose of a tensor $\mathscr{X} \in \mathbb{R}^{n \times h \times m}$, denoted by $\mathscr{X}^\top \in \mathbb{R}^{h \times n \times m}$, is obtained by transposing each frontal slice of $\mathscr{X}$ and then reversing the order of the transposed slices from the second to the $m$th. The T-inverse of a tensor $\mathscr{X} \in \mathbb{R}^{n \times n \times m}$, denoted by $\mathscr{X}^{-1} \in \mathbb{R}^{n \times n \times m}$, is defined to satisfy $\mathscr{X} \star \mathscr{X}^{-1} = \mathscr{X}^{-1} \star \mathscr{X} = \mathscr{I}$, with analogous definitions for the T-pseudoinverse. A tensor $\mathscr{X} \in \mathbb{R}^{n \times n \times m}$ is said to be T-orthogonal if it satisfies $\mathscr{X} \star \mathscr{X}^\top = \mathscr{X}^\top \star \mathscr{X} = \mathscr{I}$. All of the above operations are equivalent under the block circulant operation.  With a slight abuse of notation, we use the same superscript symbols (e.g., $\top$, $-1$) to denote both matrix and T-product-based operations.

Importantly, matrix singular value decomposition (SVD) can be extended to third-order tensors with the T-product, referred to as  tensor singular value decomposition (TSVD).

\begin{definition}
The tensor singular value decomposition (TSVD) of  a third-order tensor $\mathscr{X}\in\mathbb{R}^{n\times h\times m}$ is defined as
\begin{equation}\label{eq:tsvd}
\mathscr{X}=\mathscr{U}\star\mathscr{S}\star\mathscr{V}^{\top},
\end{equation}
where $\mathscr{U}\in\mathbb{R}^{n\times n\times m}$ and $\mathscr{V}\in\mathbb{R}^{h\times h\times m}$ are T-orthogonal, and $\mathscr{S}\in\mathbb{R}^{n\times h\times m}$ is an F-diagonal tensor (i.e., each of its frontal slices is a diagonal matrix) such that $\mathscr{S}_{jj:}\in\mathbb{R}^{m}$ are referred to the singular tubes of $\mathscr{X}$.
\end{definition}

The number of nonzero singular tubes (measured in the Frobenius norm) is referred to as the tubal rank of $\mathscr{X}$. Similar to SVD,  TSVD admits an economy-size format when $\mathscr{X}$ has low tubal rank. Suppose the tubal rank of $\mathscr{X}$ is equal to $r$ for $r\leq\min{\{n,h\}}$. Then the economy-size TSVD takes the same form as in \eqref{eq:tsvd}, with reduced factor tensors $\mathscr{U} \in \mathbb{R}^{n \times r \times m}$, $\mathscr{V} \in \mathbb{R}^{h \times r \times m}$, and $\mathscr{S} \in \mathbb{R}^{r \times r \times m}$. Moreover, truncating the singular tubes   based on their magnitudes can yield the best low-tubal-rank approximation of $\mathscr{X}$ \cite{zhang2016exact}.

TSVD can be efficiently computed using the Fourier transform combined with SVD.  A key property of block circulant matrices is that they diagonalize into block diagonal matrices in the Fourier domain. Specifically, for a tensor $\mathscr{X}\in\mathbb{R}^{n\times h\times m}$, the Fourier transform of $\texttt{bcirc}(\mathscr{X})$ gives
\begin{equation*}
    \mathcal{F}\{\texttt{bcirc}(\mathscr{X})\} = \texttt{blkdiag}(\textbf{X}_1, \textbf{X}_2,\dots,\textbf{X}_m),
\end{equation*}
where $\textbf{X}_j\in\mathbb{R}^{n\times h}$, $\mathcal{F}\{\cdot\}$ denotes the discrete Fourier transform, and \texttt{blkdiag} is the MATLAB block diagonal operator. Applying SVD to each $\mathbf{X}_j$ yields $\mathbf{X}_j = \mathbf{U}_j \mathbf{S}_j \mathbf{V}_j^\top$, and the factor tensor $\mathscr{U}$ is constructed as
\begin{equation*}
    \mathscr{U} = \texttt{un-bcirc}\Big(\mathcal{F}^{-1}\big\{\texttt{blkdiag}(\textbf{U}_1, \textbf{U}_2,\dots,\textbf{U}_m)\big\}\Big),
\end{equation*}
where \texttt{un-bcirc} denotes the reverse operation of \texttt{bcirc}. The tensors $\mathscr{S}$ and $\mathscr{V}$ can be constructed analogously. It is crucial to emphasize that the TSVD of $\mathscr{X}$ is not equivalent to the SVD of $\texttt{bcirc}(\mathscr{X})$.

Finally, tensor eigenvalue decomposition (TEVD) can be defined and computed analogously. 

\begin{definition}The
tensor eigenvalue decomposition (TEVD) of a third-order tensor $\mathscr{X}\in\mathbb{R}^{n\times n\times m}$ is defined as
\begin{equation}
\mathscr{X}=\mathscr{U}\star\mathscr{D}\star\mathscr{U}^{-1},
\end{equation}
where $\mathscr{U}\in\mathbb{C}^{n\times n\times m}$ and $\mathscr{D}\in\mathbb{C}^{n\times n\times m}$ is F-diagonal, with tubes $\mathscr{D}_{jj:} \in \mathbb{C}^{m}$ referred to as the eigentubes of $\mathscr{X}$.
\end{definition}

\subsection{Dynamic Mode Decomposition}
Dynamic mode decomposition (DMD),  introduced by Schmid in 2010 \cite{schmid2010dynamic}, is a powerful tool for analyzing high-dimensional dynamical systems using time-series measurements. The primary objective of DMD is to approximate the underlying system dynamics using a best-fit, low-dimensional linear operator that captures the temporal evolution of system states.  Given a sequence of state snapshots $\{\textbf{x}_0,\textbf{x}_1,\dots,\textbf{x}_T\}$ with $\textbf{x}_t\in\mathbb{R}^n$, DMD aims to identify a linear mapping between successive observations such that
\begin{equation}
    \textbf{x}_{t+1} = \textbf{A}\textbf{x}_t, 
\end{equation}
where $\textbf{A}\in\mathbb{R}^{n\times n}$ is \textcolor{black}{called} the state transition matrix.

To estimate $\mathbf{A}$, the state snapshots are arranged into two time-shifted matrices as
\begin{align*}
    \textbf{X}_{-} & = \begin{bmatrix}
        \textbf{x}_0 & \textbf{x}_1 & \cdots & \textbf{x}_{T-1}
    \end{bmatrix},\\
    \textbf{X}_{+} & = \begin{bmatrix}
        \textbf{x}_1 & \textbf{x}_2 & \cdots & \textbf{x}_{T}
    \end{bmatrix},
\end{align*}
such  that $\textbf{X}_{+}=\textbf{A}\textbf{X}_{-}$. Since $\textbf{A}$ may be large or ill-conditioned, DMD constructs a low-rank approximation via the economy-size SVD of $\textbf{X}_{-}$. Suppose that $\textbf{X}_{-} = \textbf{U}\textbf{S}\textbf{V}^\top$ where $\textbf{U}\in\mathbb{R}^{n\times r}$, $\textbf{S}\in\mathbb{R}^{r\times r}$, and $\textbf{V}\in\mathbb{R}^{(T-1)\times r}$ for $r\leq n$. The reduced-order approximation of the state transition matrix is obtained by projecting $\mathbf{A}$ onto the subspace spanned by $\mathbf{U}$, i.e., 
\begin{equation}
    \tilde{\textbf{A}} = \textbf{U}^\top\textbf{X}_{+}\textbf{V}\textbf{S}^{-1}\in\mathbb{R}^{r\times r}.
\end{equation}
Finally, the DMD modes are obtained by mapping the eigenvectors of $\tilde{\mathbf{A}}$ back to the full state space, expressed as
\begin{equation}
    \textbf{M} = \textbf{U}\textbf{W}\in\mathbb{R}^{n\times r},
\end{equation}
where the columns of $\mathbf{W} \in \mathbb{C}^{r \times r}$ are the eigenvectors of $\tilde{\mathbf{A}}$.

While  DMD is both elegant and computationally efficient, it is inherently limited to analyzing vector-based time-series data and cannot directly accommodate multiway (i.e., tensor-valued) observations. Flattening such tensor-structured data into vectors for use with standard DMD may disrupt the inherent multilinear structure and obscure important inter-mode dependencies. This loss of structural information can degrade interpretability, limit predictive accuracy, and compromise the discovery of meaningful spatiotemporal patterns.

\section{Tensor Dynamic Mode Decomposition}\label{sec:tdmd}
We extend the DMD framework to third-order tensors using T-product algebra, referred to as tensor dynamic mode decomposition (TDMD). Unlike traditional unfolding methods that flatten multiway data into matrices and risk losing latent correlations, TDMD operates directly in tensor space by leveraging TSVD and TEVD, leading to improved precision and efficiency. TDMD seeks to identify the underlying dynamics that govern the discrete-time evolution of a tensor-based dynamical system defined by the T-product, i.e., 
\begin{equation}
    \mathscr{X}_{t+1}=\mathscr{A}\star \mathscr{X}_t,
\end{equation}
where $\mathscr{A}\in\mathbb{R}^{n\times n\times m}$ is the state transition tensor, and $\mathscr{X}_t\in\mathbb{R}^{n\times h\times m}$ is the system state at time $t$.

Assume the state data tensors are formed by concatenating the state snapshots along the second mode, i.e., 
\begin{align*}
    \mathscr{X}_{-}&=\begin{bmatrix}
        \mathscr{X}_0 & \mathscr{X}_1 & \cdots & \mathscr{X}_{T-1}
    \end{bmatrix}\in\mathbb{R}^{n\times h(T-1)\times m},\\
\mathscr{X}_{+} &=\begin{bmatrix}
        \mathscr{X}_1 & \mathscr{X}_2 & \cdots & \mathscr{X}_{T}
    \end{bmatrix}\in\mathbb{R}^{n\times h(T-1)\times m},
\end{align*}
where $T$ is the total number of available state snapshots. Based on the properties of row block tensors under the T-product, it follows immediately that
\begin{equation}\label{eq:sys}
    \mathscr{X}_{+}=\mathscr{A}\star\mathscr{X}_{-}.
\end{equation}
The state transition tensor $\mathscr{A}$ then can be expressed as $\mathscr{A}=\mathscr{X}_{+}\star\mathscr{X}_{-}^{\dagger}$, where $\dagger$ denotes the T-pseudoinverse operation. It has been established that $\mathscr{A}$ can be uniquely identified if and only if all entries of the singular tubes  of $\mathscr{X}_{+}$ in the Fourier domain are nonzero \cite{10848219}. However, directly computing $\mathscr{A}$ is often computationally intensive, particularly for high-dimensional or large-scale systems. Our objective therefore is to efficiently compute a reduced-order approximation of  $\mathscr{A}$ that preserves the multilinear dynamics of the data and its corresponding TDMD modes.

Similar to standard DMD, we begin by computing the economy-size TSVD of the state data tensor $\mathscr{X}_{-}$, assuming the tubal rank is equal to $r$ for $r\leq n$, which yields
\begin{equation*}
    \mathscr{X}_{-}=\mathscr{U}\star\mathscr{S}\star\mathscr{V}^\top,
\end{equation*}
where $\mathscr{U}\in\mathbb{R}^{n\times r\times m}$ and $\mathscr{V}\in\mathbb{R}^{h(T-1)\times r\times m}$ are factor tensors, and $\mathscr{S}\in\mathbb{R}^{r\times r\times m}$ is an F-diagonal tensor containing the singular tubes of $\mathscr{X}_{-}$. Substituting this decomposition into \eqref{eq:sys}, we can rewrite it as 
\begin{equation*}
    \mathscr{X}_{+}=\mathscr{A}\star (\mathscr{U}\star\mathscr{S}\star\mathscr{V}^\top).
\end{equation*}
By multiplying both sides on the left by $\mathscr{U}^\top $ and on the right by
$\mathscr{V}\star\mathscr{S}^{-1}$, we obtain
\begin{equation*}
   \mathscr{U}^\top\star \mathscr{X}_{+}\star \mathscr{V}\star\mathscr{S}^{-1} = \mathscr{U}^\top\star\mathscr{A}\star \mathscr{U}.
\end{equation*}
This leads to the reduced-order state transition tensor
\begin{equation}
    \tilde{\mathscr{A}} = \mathscr{U}^\top\star \mathscr{X}_{+}\star \mathscr{V}\star\mathscr{S}^{-1}\in\mathbb{R}^{r\times r\times m}.
\end{equation}
The corresponding TDMD modes can be obtained by solving the TEVD of $\tilde{\mathscr{A}}$. Suppose the TEVD  is given by
\begin{equation*}
    \tilde{\mathscr{A}} = \mathscr{W}\star\mathscr{D}\star\mathscr{W}^{-1},
\end{equation*}
where $\mathscr{W}\in\mathbb{C}^{r\times r\times m}$ is a factor tensor, and $\mathscr{D}\in\mathbb{C}^{r\times r\times m}$ is an F-diagonal tensor containing the eigentubes of $\tilde{\mathscr{A}}$. The TDMD modes are then reconstructed as
\begin{equation}
    \mathscr{M}=\mathscr{U}\star\mathscr{W}\in\mathbb{C}^{n\times r\times m},
\end{equation}
where $\mathscr{M}_{:j:}\in\mathbb{R}^{n\times m}$ is the $j$th TDMD mode, and $\mathscr{D}_{jj:}$
is the corresponding eigentubes. 

\begin{remark}
    While the procedure of TDMD may resemble that of standard DMD, the two are fundamentally distinct in that TDMD preserves the inherent multilinear structure of the data through TSVD and TEVD, two tensor factorizations that extend beyond traditional matrix operations with novel concepts of singular tubes and eigentubes.  This structure-preserving approach enables TDMD to exploit latent multiway dependencies, leading to more accurate and efficient dynamical modeling of higher-order dynamical systems.
\end{remark}

\begin{remark}
    If the state data tensor $\mathscr{X}_{-}$ contains singular tubes with small magnitudes (i.e.,  $\|\mathscr{S}_{jj:}\|$ are small), the model order can be further reduced by truncating these insignificant singular tubes. Specifically, a low-tubal-rank approximation of $\mathscr{X}_{-}$ can be obtained by retaining only the leading $k$ singular tubes ranked by their Frobenius norms, resulting in
    \begin{equation*}
         \mathscr{X}_{-}\approx\mathscr{U}_k\star\mathscr{S}_k\star\mathscr{V}_k^\top,
    \end{equation*}
    where $\mathscr{U}_k \in \mathbb{R}^{n \times k \times m}$ and $\mathscr{V}_k \in \mathbb{R}^{h(T-1) \times k \times m}$ are truncated tensors, and $\mathscr{S}_k \in \mathbb{R}^{k \times k \times m}$ is an F-diagonal tensor containing the dominant singular tubes. The subsequent steps of  TDMD  remain unchanged. The resulting reduced-order state transition tensor  becomes $\tilde{\mathscr{A}} \in \mathbb{R}^{k \times k \times m}$ with the corresponding TDMD modes $\mathscr{M}\in\mathbb{C}^{n\times k\times m}$. Suppose  we truncate $l=n-k$  singular tubes in the TSVD of $\mathscr{X}_{-}$. The total number of parameters in the reduced-order  system using TDMD is $(n-l)^2m$. In contrast, for the same level truncation (i.e., truncating $l$ singular values), the total number
of parameters in the reduced-order system using DMD with data flattening is $(nm-l)^2$. 
\end{remark}

\begin{remark}
    TDMD can be efficiently implemented in the Fourier domain. Let $\textbf{X}^{(j)}_{-}$ and $\textbf{X}^{(j)}_{+}$ denote the diagonal block matrices of $\texttt{bcirc}(\mathscr{X}_{-})$ and $\texttt{bcirc}(\mathscr{X}_{+})$ in the Fourier domain for $j=1,2,\dots,m$. Suppose that the matrix SVDs of $\textbf{X}^{(j)}_{-}$ are expressed as $\textbf{X}^{(j)}_{-}=\textbf{U}_j\textbf{S}_j\textbf{V}_j^\top$. Then the reduced-order state transition tensor can be computed as
    \begin{equation*}
        \tilde{\mathscr{A}} = \texttt{un-bcirc}\Big(\mathcal{F}^{-1}\big\{\texttt{blkdiag}(\tilde{\textbf{A}}_1, \tilde{\textbf{A}}_2,\dots,\tilde{\textbf{A}}_m)\big\}\Big),
    \end{equation*}
    where 
    \begin{equation*}
        \tilde{\textbf{A}}_j = \textbf{U}_j^\top \textbf{X}_{+}^{(j)}\textbf{V}_j\textbf{S}_j^{-1}\in\mathbb{R}^{r\times r}
    \end{equation*}
    for $j=1,2,\dots,m$. The corresponding TDMD modes can be constructed similarly by computing the matrix EVDs of $\tilde{\textbf{A}}_j$.  Detailed steps of  TDMD  are summarized in Algorithm 1. This slice-wise formulation significantly enhances the computational efficiency, yielding an estimated time complexity of $\mathcal{O}(nThm\log{m}+nThms)$ where $s=\min{\{n,T\}}$. In comparison, standard DMD with data flattening incurs a higher time complexity of $\mathcal{O}(n^2m^2+nThm^3s)$. 
\end{remark}

 \begin{algorithm}[t]
\caption{Tensor dynamic mode decomposition}\label{alg:tdmd}
\begin{algorithmic}[1]
\State \textbf{Input:} Given the state data tensors $\mathscr{X}_{-}$ and $\mathscr{X}_{+}$ 
\State Compute the diagonal block matrices $\textbf{X}^{(j)}_{-}$ and $\textbf{X}^{(j)}_{+}$  of $\texttt{bcirc}(\mathscr{X}_{-})$ and $\texttt{bcirc}(\mathscr{X}_{+})$ in the Fourier domain
\For{$j=1,2,\dots,m$}
\State Compute the economy-size  SVDs of $\textbf{X}^{(j)}_{-}$, i.e.,  
$
    \textbf{X}^{(j)}_{-}=\textbf{U}_{j}\textbf{S}_{j}\textbf{V}_{j}^\top
$
\State Set $\tilde{\textbf{A}}_j = \textbf{U}_j^\top \textbf{X}_{+}^{(j)}\textbf{V}_j\textbf{S}_j^{-1}$
\State Compute the  EVDs of $\tilde{\textbf{A}}_j$, i.e.,
$
    \tilde{\textbf{A}}_j=\textbf{W}_j\textbf{D}_j\textbf{W}_j^{-1}
$
\EndFor
\State The reduced-order state transition tensor is computed as
\begin{equation*}
    \tilde{\mathscr{A}} = \texttt{un-bcirc}\Big(\mathcal{F}^{-1}\big\{\texttt{blkdiag}(\tilde{\textbf{A}}_1, \tilde{\textbf{A}}_2,\dots,\tilde{\textbf{A}}_m)\big\}\Big)
\end{equation*}
\State The TDMD modes are computed as
\begin{equation*}
    \mathscr{M} = \texttt{un-bcirc}\Big(\mathcal{F}^{-1}\big\{\texttt{blkdiag}(\textbf{M}_1, \textbf{M}_2,\dots,\textbf{M}_m)\big\}\Big),
\end{equation*}
where $\textbf{M}_j=\textbf{U}_j\textbf{W}_j$
\State \textbf{Output:} Reduced-order state transition tensor $\tilde{\mathscr{A}}$ and TDMD modes $\mathscr{M}$.
\end{algorithmic}
\end{algorithm}

DMD modes provide a powerful data-driven representation of coherent spatiotemporal structures that govern the evolution of complex dynamical systems. Building on this foundation, TDMD modes extend DMD to multiway data by preserving and leveraging the multilinear structure inherent in higher-order dynamical systems. Instead of flattening data into matrices, TDMD operates directly on tensorial representations, enabling the identification of multidimensional modes that capture interactions across multiple axes (e.g., space, time, and channel). This richer structure allows for a more accurate and interpretable decomposition of complex dynamics, particularly in systems with naturally multi-modal data, such as video, neuroimaging, and higher-order networks. In the following, we demonstrate two key applications of TDMD: state reconstruction and dynamic component separation.

\subsection{State Reconstruction}
An immediate application of TDMD is the reconstruction of state trajectories from low-rank dynamic representations. Once the TDMD modes and the corresponding eigentubes have been computed, the time evolution of the system can be approximated in the low-dimensional coordinate system. First, we express the initial snapshot as the T-product between the TDMD modes $\mathscr{M}\in\mathbb{R}^{n\times r\times m}$ and a  tensor of modal amplitudes $\mathscr{B}\in\mathbb{R}^{r\times h\times m}$ such that
\begin{equation*}
    \mathscr{X}_0 = \mathscr{M}\star \mathscr{B}.
\end{equation*}
\textcolor{black}{Once the initial modal amplitude tensor  $\mathscr{B}=\mathscr{X}_0\star \mathscr{M}^{\dagger}$ is computed}, the system state at time $t$ can be reconstructed as
\begin{equation*}
    \mathscr{X}_t = \mathscr{M}\star \mathscr{D}^{t}\star\mathscr{B},
\end{equation*}
where \textcolor{black}{$\mathscr{D}^t\in\mathbb{R}^{r\times r\times m}$ denotes the $t$th T-product power of the F-diagonal tensor containing the eigentubes of $\tilde{\mathscr{A}}$.} In addition to reconstruction, TDMD even can predict future states of the system outside the observed time window. 

\subsection{Dynamic Component Separation}
TDMD can be employed to separate distinct dynamic components within  multiway data. In particular, a sequence of snapshots can be decomposed into persistent  and transient  dynamics based on the magnitudes of the eigentubes. \textcolor{black}{Similar to DMD, modes with eigentubes of magnitude close to one typically represent persistent, long-term behaviors, while those with significantly smaller magnitudes correspond to transient components that decay over time.} Such separation is especially valuable in applications like video processing, biomedical signal analysis, and fluid dynamics, where identifying dominant or ephemeral behaviors can lead to more effective modeling, prediction, and control strategies.

To achieve this decomposition, we first classify the eigentubes $\mathscr{D}_{jj:}\in\mathbb{R}^{m}$ into two categories based on their Frobenius norms:  those with $\|\mathscr{D}_{jj:}\|> \epsilon$ are considered persistent, while those satisfying $\|\mathscr{D}_{jj:}\|\leq \epsilon$ are considered transient, where $\epsilon$ is a predefined threshold chosen close to one (e.g., 0.995). Let $\mathscr{D}_{\text{pers}}$ and $\mathscr{D}_{\text{trans}}$ denote the F-diagonal tensors composed of the eigentubes in the persistent and transient categories, respectively. The corresponding dynamic components at time $t$ can then be reconstructed as
\begin{equation*}
    \mathscr{X}_{\text{pers}}(t) = \mathscr{M}\star \mathscr{D}_{\text{pers}}^{t}\star\mathscr{B},\quad
    \mathscr{X}_{\text{trans}}(t) = \mathscr{M}\star \mathscr{D}_{\text{trans}}^{t}\star\mathscr{B},
\end{equation*}
where $\mathscr{B}$ is the initial modal amplitude tensor defined previously. The full state at time $t$ therefore can be expressed as the sum of the persistent and transient components: $\mathscr{X}_t = \mathscr{X}_{\text{pers}}(t)+ \mathscr{X}_{\text{trans}}(t)$.

In summary, TDMD not only offers a compact and low-rank representation of multiway dynamical data but also facilitates interpretable temporal decomposition by enabling the separation of long-term trends from short-lived, transient phenomena. This capability provides a principled framework for analyzing complex spatiotemporal systems across a variety of applications. Moreover, all computations involved in state reconstruction and dynamic component separation can be performed efficiently in the Fourier domain through the slice-wise implementation.

\section{Numerical Examples}\label{sec:num}
All numerical experiments were conducted using MATLAB R2024b on a machine equipped with an M1 Pro CPU and 16 GB of memory, utilizing the tensor-tensor-product-toolbox-master \cite{lu2018tensor}. The code used for these experiments is available at \url{https://github.com/ZiqinHe/Tensor_DMD}.

\begin{figure}[t]
    \centering
    \includegraphics[width=\linewidth]{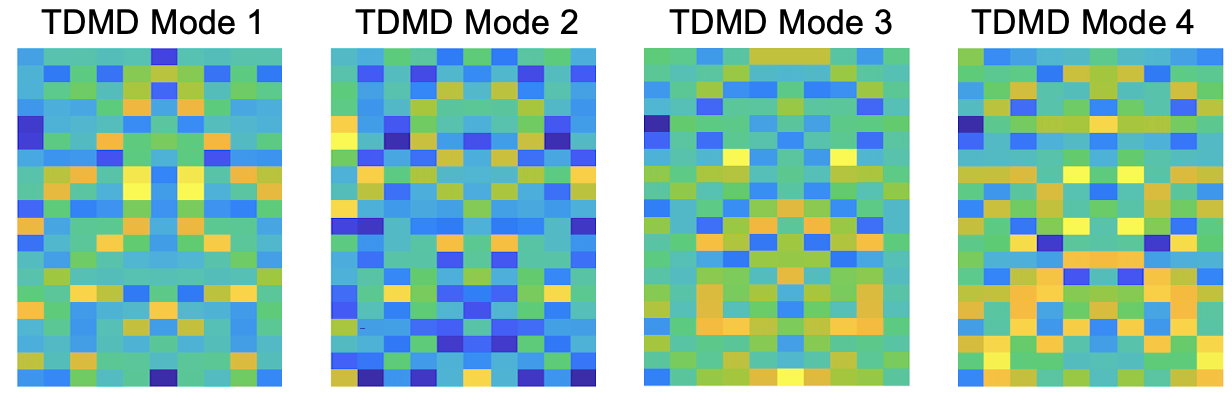}
    \caption{First four TDMD modes of the synthetically generated data.}
    \label{fig:modes}
\end{figure}

\subsection{Synthetic Data}
In this example, we evaluated the performance of TDMD with synthetic data. Specifically, we generated a sequence of random states $\mathscr{X}_t \in \mathbb{R}^{10 \times 1 \times 6}$ for $t = 1, 2, \dots, 20$. Using these snapshots, we constructed the two input tensors $\mathscr{X}_{-}$ and $\mathscr{X}_{+}$, corresponding to lagged state pairs. Applying TDMD to this data, we extracted the TDMD modes  (Fig. \ref{fig:modes}), which capture the dominant spatiotemporal structures of this multiway data.  We further assessed the reconstruction quality using the extracted TDMD modes by reconstructing the original state sequence. The performance was compared against standard DMD with data flattening in terms of computational time, memory usage, and total relative reconstruction error. The memory usage measures the size of the identified reduced-order system, and the total relative reconstruction error is defined as
$
    \sum_{t=1}^{20}\|\hat{\mathscr{X}}_t-\mathscr{X}_t\|/\|\mathscr{X}_t\|
$, where $\hat{\mathscr{X}}_t$ is the reconstructed state at time $t$. The quantitative comparison of TDMD and standard DMD across different singular tubes/values truncation levels is summarized in Table \ref{tab:tdmd-unfold}. Notably, TDMD consistently outperforms standard DMD in both computational efficiency and memory usage. More importantly, it achieves lower relative reconstruction errors across all settings. These results highlight the superior accuracy and efficiency of TDMD in modeling the dynamics of multiway data.

\begin{table*}[t]
\centering
\caption{Comparison of computational time, memory usage, and total relative reconstruction error between TDMD and  DMD with data flattening across different singular tube/value truncation levels. \textcolor{black}{Note that while the total reconstruction errors for DMD are 1.3527 across truncation levels, they are not identical if including more decimal places.}}
\renewcommand{\arraystretch}{1.2}
\setlength{\tabcolsep}{12pt}
\small
\begin{tabular}{c c c c c c c}
\hline
\multirow{2}{*}{Truncation Level} & \multicolumn{2}{c}{Time (seconds)} & \multicolumn{2}{c}{Memory Usage (MB)} & \multicolumn{2}{c}{Total Relative Error} \\  
& TDMD &  DMD & TDMD &  DMD & TDMD &  DMD \\  
\hline
0  & 0.0449 & 0.2049 & 0.5784 & 55.3596 & $5.15 \times 10^{-11}$ & $1.3527$ \\
1  & 0.0534 & 0.1895 & 0.5191 & 49.6857 & $7.03 \times 10^{-10}$ & $1.3527$ \\
2  & 0.0581 & 0.1594 & 0.4630 & 44.3184 & $5.77 \times 10^{-9}$ & $1.3527$ \\
3  & 0.0438 & 0.1434 & 0.4102 & 39.2578 & $3.27 \times 10^{-8}$ & $1.3527$ \\
4  & 0.0413 & 0.1190 & 0.3605 & 34.5040 & $1.53 \times 10^{-7}$ & $1.3527$ \\
5  & 0.0401 & 0.0960 & 0.3140 & 30.0568 & $6.49 \times 10^{-7}$ & $1.3527$ \\
6  & 0.0375 & 0.0853 & 0.2708 & 25.9163 & $2.70 \times 10^{-6}$ & $1.3527$ \\
\hline
\end{tabular}
\label{tab:tdmd-unfold}
\end{table*}

\subsection{A Case Study on Video Data}
\begin{figure}
    \centering
    \includegraphics[width=\linewidth]{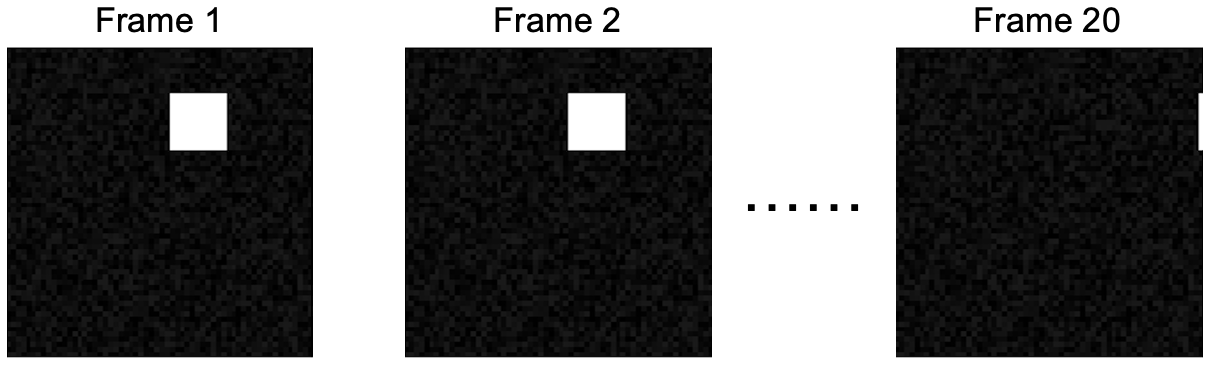}
    \caption{A video dataset consisting of 20 frames, featuring a stationary, noise-corrupted background with a single white square moving horizontally from left to right at a constant speed.}
    \label{fig:video}
\end{figure}

\begin{figure}
    \centering
    \includegraphics[width=0.8\linewidth]{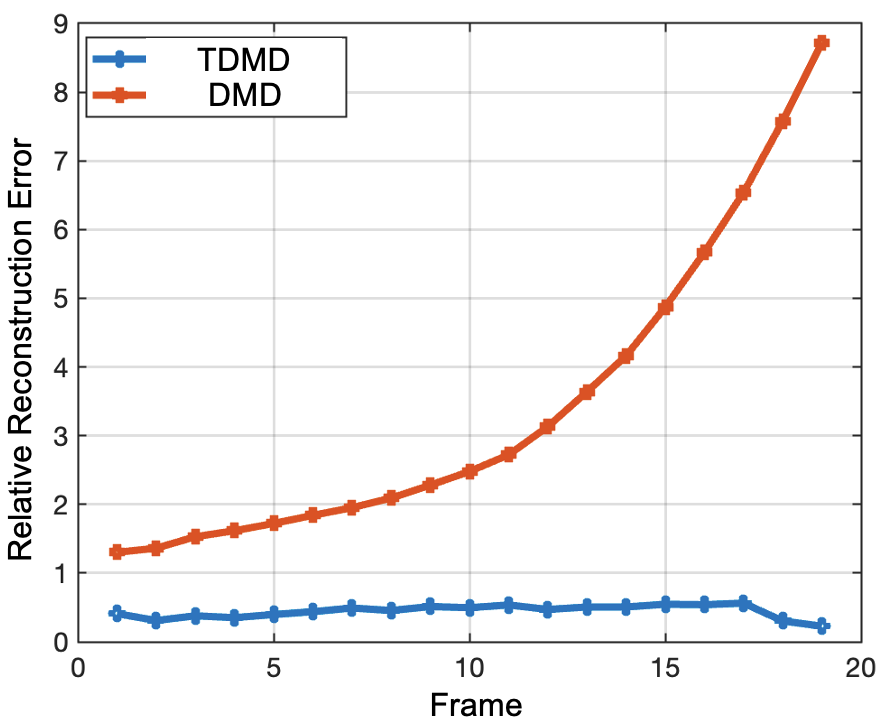}
    \caption{Comparison of relative reconstruction errors between TDMD and standard DMD at each frame on the video dataset.}
    \label{fig:error}
\end{figure}

\begin{figure}
    \centering
    \includegraphics[width=0.85\linewidth]{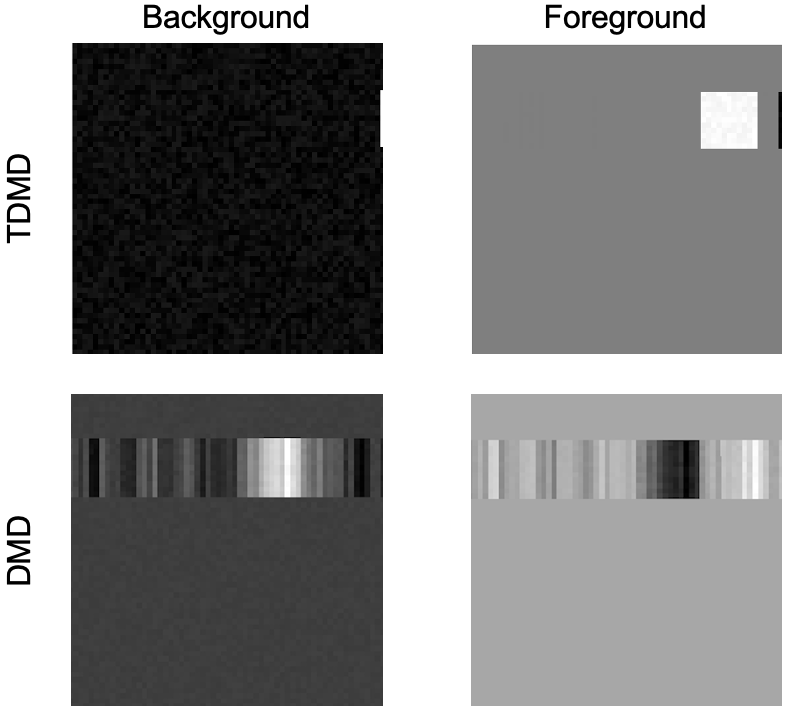}
    \caption{Comparison of background and foreground separation on Frame 10 using TDMD and standard DMD.}
    \label{fig:separation}
\end{figure}

In this example, we applied TDMD to a video dataset to separate dynamic foreground elements from the static background. The video features a stationary, noise-corrupted background with a single white square moving horizontally from left to right at a constant speed (Fig. \ref{fig:video}). The dataset contains 20 frames, each captured at a distinct time point and having a resolution of $60 \times 60$ pixels. \textcolor{black}{We first assessed the performance of recovering each frame (except Frame 1)  by comparing TDMD and standard DMD with data flattening in terms of relative reconstruction error across all frames (without any truncation).} As shown in Fig. \ref{fig:error}, TDMD consistently achieves significantly lower reconstruction errors for each frame, demonstrating its superior ability to preserve and exploit the spatiotemporal structure of the data. Beyond reconstruction, we further applied TDMD for dynamic component separation. Notably, TDMD effectively isolates the static, noise-corrupted background from the moving white square in the foreground (Fig. \ref{fig:separation}), clearly highlighting its ability to distinguish persistent spatial features from time-varying dynamics. In contrast, standard DMD fails to achieve such a clean separation. These findings highlight  the advantage of TDMD in accurately modeling and disentangling complex spatiotemporal phenomena in multiway data, offering a powerful tool for video analysis and beyond.

\section{Conclusion}\label{sec:con}
In this letter, we proposed tensor dynamic mode decomposition (TDMD), a novel extension of DMD to third-order tensors. Fundamentally distinct from standard DMD, TDMD leverages tensor  factorization techniques including TSVD and TEVD to construct a reduced-order state transition tensor and extract TDMD modes that preserve the multilinear structure of the data. This structure-preserving framework enables TDMD to effectively perform tasks such as state reconstruction and dynamic component separation. Numerical experiments highlight the superiority of TDMD over standard DMD in terms of computational efficiency, memory consumption, and reconstruction accuracy. \textcolor{black}{Last but not least, the framework can be readily generalized to higher-order tensors through the generalized T-product and its associated algebra.} 

Future work will focus on extending TDMD to handle higher-order tensors beyond third order, enabling analysis of even more complex multiway data structures. Additionally, incorporating noise robustness and regularization into the TDMD framework could improve performance in real-world settings where data is often corrupted or incomplete. Another promising direction is integrating TDMD with machine learning models to enable hybrid approaches that combine interpretable dynamic modeling with predictive power.  Finally, we plan to apply the proposed framework to a broad range of real-world multiway datasets to validate its effectiveness and tackle practical challenges such as noise, missing data, and complex dynamic interactions intrinsic to real-world systems.

\section*{References}
\bibliographystyle{IEEEtran}
\bibliography{reference.bib}
\end{document}